\documentclass[pdflatex,sn-mathphys-num]{sn-jnl}

\usepackage{graphicx}%
\usepackage{multirow}%
\usepackage{amsmath,amssymb,amsfonts}%
\usepackage{amsthm}%
\usepackage{mathrsfs}%
\usepackage[title]{appendix}%
\usepackage{xcolor}%
\usepackage{textcomp}%
\usepackage{manyfoot}%
\usepackage{booktabs}%
\usepackage{algorithm}%
\usepackage{algorithmicx}%
\usepackage{algpseudocode}%
\usepackage{listings}%
\usepackage{lscape}

\newcommand{\bbeta}{{\mbox{\boldmath $\beta$}}}

\newcommand{\bmu}{{\mbox{\boldmath $\mu$}}}
\newcommand{\bSigma}{{\mbox{\boldmath $\Sigma$}}}

\newcommand{\ba}{{\mbox{\boldmath $a$}}}

\newcommand{\x}{{\mbox{\boldmath $x$}}}

\newcommand{\X}{{\mbox{\boldmath $X$}}}
\newcommand{\Y}{{\mbox{\boldmath $Y$}}}

\newcommand{\W}{{\mbox{\boldmath $W$}}}

\theoremstyle{thmstyleone}%
\theoremstyle{thmstyletwo}%
\newtheorem{remark}{Remark}%

\theoremstyle{thmstylethree}%

\begin{document}

\title[Article Title]{Improving operating characteristics of clinical trials by augmenting control arm using propensity score-weighted borrowing-by-parts power prior}


\author*[1]{\fnm{Apu Chandra} \sur{Das}}\email{apdas@unmc.edu}

\author[2]{\fnm{Sakib} \sur{Salam}}\email{msalam@mcw.edu}

\author[3]{\fnm{Aninda} \sur{Roy}}\email{aroy@mcw.edu}

\author[4]{\fnm{Rakhi} \sur{Chowdhury}}\email{rachowdhury@unmc.edu}

\author[5]{\fnm{Antar Chandra} \sur{Das}}\email{antar2305341489@diu.edu.bd}
\equalcont{These authors contributed equally to this work.}

\author[6]{\fnm{Ashim Chandra} \sur{Das}}\email{ashim2305341488@diu.edu.bd}
\equalcont{These authors contributed equally to this work.}

\author[7]{for the Alzheimer’s Disease Neuroimaging Initiative}

\affil*[1]{\orgdiv{Department of Biostatistics}, \orgname{University of Nebraska Medical Center}, \orgaddress{\street{984375 Nebraska Medical Center}, \city{Omaha}, \postcode{68198}, \state{Nebraska}, \country{USA}}}

\affil[2, 3]{\orgdiv{Division of Biostatistics}, \orgname{Medical College of Wisconsin}, \orgaddress{\street{8701 Watertown Plank Rd}, \city{Milwaukee}, \postcode{53226}, \state{Wisconsin}, \country{USA}}}

\affil[4]{\orgdiv{Department of Pharmaceutical Sciences}, \orgname{University of Nebraska Medical Center}, \orgaddress{\street{984375 Nebraska Medical Center}, \city{Omaha}, \postcode{68198}, \state{Nebraska}, \country{USA}}}

\affil[5,6]{\orgdiv{Department of Software Engineering}, \orgname{Daffodil International  University}, \orgaddress{\street{Daffodil Smart City}, \city{Savar}, \postcode{1216}, \state{Dhaka}, \country{Bangladesh}}}

\affil[7]{\orgdiv{Data used in preparation of this article were obtained from the Alzheimer’s Disease Neuroimaging Initiative (ADNI) database (adni.loni.usc.edu). As such, the investigators within the ADNI contributed to the design and implementation of ADNI and/or provided data but did not participate in analysis or writing of this report. A complete listing of ADNI investigators can be found at: http://adni.loni.usc.edu/wp-content/uploads/how\_to\_apply/ADNI\_Acknowledgement\_List.pdf}}


\abstract{Borrowing external data can improve estimation efficiency but may introduce bias when populations differ in covariate distributions or outcome variability. A proper balance needs to be maintained between the two datasets to justify the borrowing. We propose a propensity score weighting borrowing-by-parts power prior (PSW-BPP) that integrates causal covariate adjustment through propensity score weighting with a flexible Bayesian borrowing approach to address these challenges in a unified framework. The proposed approach first applies propensity score weighting to align the covariate distribution of the external data with that of the current study, thereby targeting a common estimand and reducing confounding due to population heterogeneity. The weighted external likelihood is then incorporated into a Bayesian model through a borrowing-by-parts power prior, which allows distinct power parameters for the mean and variance components of the likelihood, enabling differential and calibrated information borrowing. Additionally, we adopt the idea of the minimal plausibility index (mPI) to calculate the power parameters. This separate borrowing provides greater robustness to prior–data conflict compared with traditional power prior methods that impose a single borrowing parameter. We study the operating characteristics of PSW-BPP through extensive simulation and a real data example. Simulation studies demonstrate that PSW-BPP yields more efficient and stable estimation than no borrowing and fixed borrowing, particularly under moderate covariate imbalance and outcome heterogeneity. The proposed framework offers a principled and extensible methodological contribution for Bayesian inference with external data in observational and hybrid study designs.}


\maketitle

\section{Introduction}\label{intro}
In clinical trials, randomized controlled trials (RCTs) are considered to be the gold standard to evaluate the effectiveness of a treatment during the process of drug development \cite{bakker2023contribution}. Despite their utility in producing reliable and unbiased estimates of treatment efficacy, RCTs often present practical challenges. These include delayed access to promising therapies, lengthy timelines, potential exposure of participants to suboptimal treatments, and logistical difficulties, particularly in pediatric populations \cite{purpura2022role}. In recent decades, growing attention has been directed toward integrating real-world data (RWD) into the drug development process to enhance both the efficiency and ethical conduct of clinical trials \cite{kidwell2022application}. One promising approach involves using external clinical trial data to augment the control arm of new studies. When incorporated appropriately, such external control data can reduce the number of patients exposed to less effective treatments, decrease the size of the control group, and improve the overall feasibility and ethical balance of trials \cite{ghadessi2020roadmap}. Although this technique can be applied in various fields—such as clinical trials, genetics, healthcare, manufacturing, environmental health, engineering, economics, and business—it may be particularly useful in pediatric clinical trials, where patient recruitment is more challenging due to scheduling difficulties with children and parents, the need for age-appropriate formulations, concerns about palatability, and ethical constraints \cite{boat2012safe}.

It is important to assess systematic differences between the current study population and subjects from external studies before leveraging RWD to minimize the risk of potential biases. The propensity score (PS) \cite{rosenbaum1983central} is a widely used statistical tool for mitigating such biases. In addition to PS-based approaches, several Bayesian methods have been proposed to incorporate external information as informative priors. These include the power prior \cite{ibrahim2015power}, the commensurate power prior \cite{hobbs2011hierarchical}, the meta-analytic-predictive (MAP) prior \cite{schmidli2014robust}, the elastic MAP prior \cite{zhang2021elastic}, and the scale-transformed power prior \cite{alt2023scale}.
In pediatric clinical trials, where borrowing adult data is common due to limited pediatric data, PS-integrated Bayesian methods have been used effectively. PS stratification helps reduce systematic differences, making the two datasets more comparable \cite{lunceford2004stratification, rosenbaum1984reducing}. A recent study \cite{lin2019propensity} applied PS matching to obtain a one-to-one matching between adult and pediatric subjects, followed by the use of Bayesian borrowing approaches, such as the mixture prior and power prior, to estimate the treatment effect. Wang et al. \cite{wang2019propensity} and Lu et al. \cite{lu2022propensity} integrated a Bayesian power prior with PS stratification to derive stratum-specific posterior distributions. However, the traditional power prior uses a single power parameter uniformly across the entire external dataset, limiting its flexibility to selectively borrow from more relevant or higher-quality portions. This can lead to suboptimal borrowing, especially when external data are heterogeneous. To address this, Baron et al. \cite{baron2024enhancing} recently extended the PS-integrated power prior framework proposed by Wang and Lu \cite{wang2019propensity, lu2022propensity} to a borrowing-by-parts power prior to leverage the external data in estimating the treatment effect. In rare disease or pediatric trials with extremely small sample sizes, it may not be feasible to form multiple strata. It is also reasonable to determine the proportion of information contributed by each external subjects. In such cases, Li et al. \cite{li2025using} proposed the use of PS weighting along with Bayesian power prior to improve the operating characteristics of the power prior without requiring stratification, thereby facilitating more effective augmentation of the control arm.


The PS-integrated power prior approaches proposed by  Wang et al. \cite{wang2019propensity} and Lu et al. \cite{lu2022propensity} require an involved numerical integration to calculate an overlapping area between the two datasets, which is later used to compute power parameters. However, both methods rely on a pre-specified ``nominal number of external subjects to be borrowed $(A)$'', which may not always be practical. 
We extend the previous approaches in three ways. First, we utilize the idea of PS-weighting \cite{li2025using} instead of PS stratification. One key advantage of our weighting method—compared to the two previous approaches—is its simpler implementation. Second, we utilize the idea of borrowing-by-parts power prior \cite{yuan2022bayesian} instead of power prior to that allows us a flexible borrowing through different components of the data. Third, we incorporate the concept of the minimal plausibility index (mPI), as proposed by Baron et al. \cite{baron2024enhancing}, to calculate the power parameters.

Our weighting scheme differs from that proposed by Lin et al. \cite{lin2019propensity}. While Lin et al. \cite{lin2019propensity} utilize PS matching to ``bring in" external subjects for augmenting the current study and then use the PS as the power parameter in a subject-specific power prior, we instead assign subject-specific weights directly to each external control through its likelihood. In contrast to the approach by Li et al. \cite{li2025using}, who define weights as the PS odds, which can exceed 1, we propose using the minimum of 1 and the PS odds to ensure all weights fall within the [0, 1] range.


The rest of the article is structured as follows. Section \ref{sec:method} outlines the propensity score framework, the borrowing-by-parts power prior, and our proposed approach. Section \ref{sec:sim_study} reports the simulation settings and results used to assess the performance of the proposed method. The findings from the real data application are summarized in Section \ref{sec:case_study}. Finally, Section \ref{sec:discussion} concludes with a discussion of the proposed method and potential scopes for future research.

\section{Method} \label{sec:method}

\subsection{Study Setting} \label{sec:study_sett}

We consider a two-arm randomized controlled trial designed to evaluate the effect of an investigating treatment. Alongside the trial data, covariates associated with the outcome are available from external control patients and may be incorporated into the analysis. The primary aim is to leverage these external real-world data (RWD) to produce an unbiased and efficient estimate of the treatment effect in the current study. Throughout this paper, we refer to the external real-world controls as group $e$. Within the trial, patients randomized to control form group $c$, while those assigned to the investigating treatment form group $t$. Together, groups $c$ and $t$ comprise the trial population. For clarity, we focus on a single external dataset, though the proposed strategy can naturally be extended to settings with multiple external sources.

Let $Y_{ti} \ \text{for} \ i=1,\dots,n_t$, $Y_{cj} \ \text{for} \ j=1,\dots,n_c$, and $Y_{ek} \ \text{for} \ k=1,\dots,n_e$ be the continuous responses for the investigating treatment, concurrent control, and external control samples, respectively. Also let, $\X_{ti}, \X_{cj}, \ \text{and} \ \X_{ek}$ denote a $p$-dimensional vectors of covariates with corresponding regression coefficient $\bbeta=(\beta_1, \beta_2, \dots, \beta_p)^\prime$. Also assume $Y_{ti} \overset{\text{iid}}{\sim} {\cal N}(\theta_{t}, \sigma^2_t), Y_{cj} \overset{\text{iid}}{\sim} {\cal N} (\theta_{c}, \sigma^2_c), \ \text{and} \ Y_{ek} \overset{\text{iid}}{\sim} {\cal N} (\theta_{e}, \sigma^2_e)$, where $\theta_t, \theta_c, \ \text{and} \ \theta_e$ are the mean responses of the three groups, respectively, and $\sigma^2_t, \sigma^2_c, \ \text{and} \ \sigma^2_e$ are the variances of the three groups, respectively. The treatment effect $\theta=\theta_t-\theta_c$, which is the difference of mean responses in the treatment and control groups, is a part of $\bbeta$. We denote $D_t=\{n_t, Y_{ti}, \X_{ti}, i=1,\dots, n_t\}$, $D_c=\{n_c, Y_{cj}, \X_{cj}, j=1,\dots, n_c\}$,  and $D_e=\{n_e, Y_{ek}, \X_{ek}, e=1,\dots, n_k\}$ as investigating treatment data, concurrent control data, and external control data, respectively.

\subsection{Propensity Score} \label{sec:prop_score}
The idea of PS was first introduced by Rosenbaum and Rubin \cite{rosenbaum1983central} as a means of conducting causal inference in observational studies, and it has since been widely applied across diverse research areas \cite{mccaffrey2013tutorial, pan2018propensity, austin2021applying, d2007propensity}. In observational settings where RCTs are not feasible due to ethical, logistical, or financial reasons, treatment assignment is often influenced by patient characteristics, leading to potential confounding. The PS, defined as the probability of receiving treatment given observed covariates, serves as a balancing tool to mitigate this issue. By aligning treated and control patients with similar PS values through methods such as matching, stratification, weighting, or covariate adjustment, it is possible to reduce selection bias and approximate the balance achieved in an RCT \cite{akmal2022propensity}. This, in turn, enhances the validity and reliability of estimated causal relationships drawn from RWD. Given the initial concept, we utilize PS to balance covariates, including prognostic factors, baseline disease severity, biomarkers, and previous treatment between the current and external study populations. 

The PS, $e(\X)$, for a subject with a set of baseline covariates $\X$ can be written as

\begin{equation}
    e(\X)=P(Z=1 \mid \X),
\end{equation}
where $Z=1$ if the subject belongs to the current study and $Z=0$ if not. It denotes the probability, conditional on baseline covariates $X$, that a patient belongs to the current population instead of the external population. While logistic regression is the most commonly employed method for estimating the propensity score, nonparametric approaches can also be used. After obtaining the estimated scores, a weight is then calculated for each external subject.

\subsection{Borrowing-by-Parts Power Prior (BPP)} \label{sec:bbppp}
The power prior, introduced by Ibrahim and Chen \cite{ibrahim2000power}, is a Bayesian hierarchical model for incorporating external control data into the analysis of current studies by explicitly controlling the level of borrowing. It is determined by a tuning parameter called the power parameter, which is raised to the external data likelihood, allowing integration of informative prior knowledge. This flexibility makes the power prior particularly useful in settings with small current samples, where leveraging data can increase effective sample size and improve statistical power for decision-making.

Borrowing-by-parts power prior (BPP) \cite{yuan2022bayesian} is an extension of the power prior that allows information from external data to be incorporated selectively rather than uniformly. Instead of applying a single power parameter to the entire dataset, this approach decomposes the parameters into different components and assigns separate levels of borrowing to each part. By doing so, it provides flexibility to borrow more information where the external and current data are similar, and less where they differ. This selective borrowing enhances the relevance of external information, leading to more accurate and reliable statistical inferences.

Assuming both control groups share the same parameters, we construct the power prior by writing $\theta_c=\theta_e$ and $\sigma^2_c=\sigma^2_e$. Let $\bar Y_c \ \text{and} \ \bar Y_e$ denote the sample means, and $S^2_c \ \text{and} \ S^2_e$ denote the sample variances, of the responses in the current control group and the external control group, respectively. Then, the BPP for $\theta_c \ \text{and} \ \sigma^2_c$ can be written as

\begin{equation}
    \pi(\theta_c, \sigma^2_c \mid D_e, \ba) \propto f(\bar Y_e \mid \theta_c, \sigma^2_c)^{a_1} \times f(S^2_e \mid \sigma^2_c)^{a_2} \times \pi_0(\theta_c, \sigma^2_c), \label{eqn:base_bpp}
\end{equation}
where $f(\bar Y_e \mid \theta_c, \sigma^2_c)$ and $f(S^2_e \mid \sigma^2_c)$ are the  two different parts of the BPP corresponding to $\theta_c$ and $\sigma^2_c$, $\pi_0(\theta_c, \sigma^2_c)$ is an initial prior for $\theta_c$ and $\sigma^2_c$, and $\ba=(a_1, a_2)^\prime (0\leq a_1,a_2 \leq1)$ are the power parameters in the regression coefficient $\theta_c$ and variance $\sigma^2$, respectively.
We can complete the equation \eqref{eqn:base_bpp} as

\begin{align}
    \pi(\theta_c, \sigma^2_c \mid D_e, \ba) 
    & \propto \left \{(\sigma^2_c)^{-\frac{1}{2}} \exp \left(-\frac{(\theta_c-\hat \theta_e)^\prime (\X_e^\prime \X_e)(\theta_c-\hat\theta_e)}{2\sigma^2_c} \right) \right \}^{a_1} \nonumber\\
    & \; \; \; \;\ \; \times \left\{ (S^2_e)^{\frac{n_e-3}{2}} (\sigma^2)^{-\frac{n_e-1}{2}}\exp \left( -\frac{(n_e-1)S^2_e}{2\sigma^2_c} \right) \right\}^{a_2} \left(\frac{1}{\sigma^2_c}\right), \label{eqn:bpp}
\end{align}
where $S^2_e =\frac{1}{n_e-1} (\Y_e-\X_e\hat \theta_e)^\prime (\Y_e-\X_e\hat \theta_e)$, $\pi_0(\theta_c, \sigma^2_c)=1/\sigma^2_c$, and $\hat\theta_e$ is an estimate of treatment effect using external controls only.

\begin{remark}
    Equation \eqref{eqn:bpp} is more flexible in controlling the amount of borrowing by the different decomposed parts of the data, controlled by $a_1 \ \text{and} \ a_2$, compared to the traditional power prior \cite{ibrahim2000power}. In the case of known $\sigma^2_c$, the equation \eqref{eqn:bpp} can be reduced to the original power prior.
\end{remark}

\subsection{Propensity Score-Weighted Borrowing-by-parts Power Prior (PSW-BPP)} \label{sec:ps_bbppp}
\begin{figure} [ht]
\centering
\includegraphics[width=14cm,height=8cm]{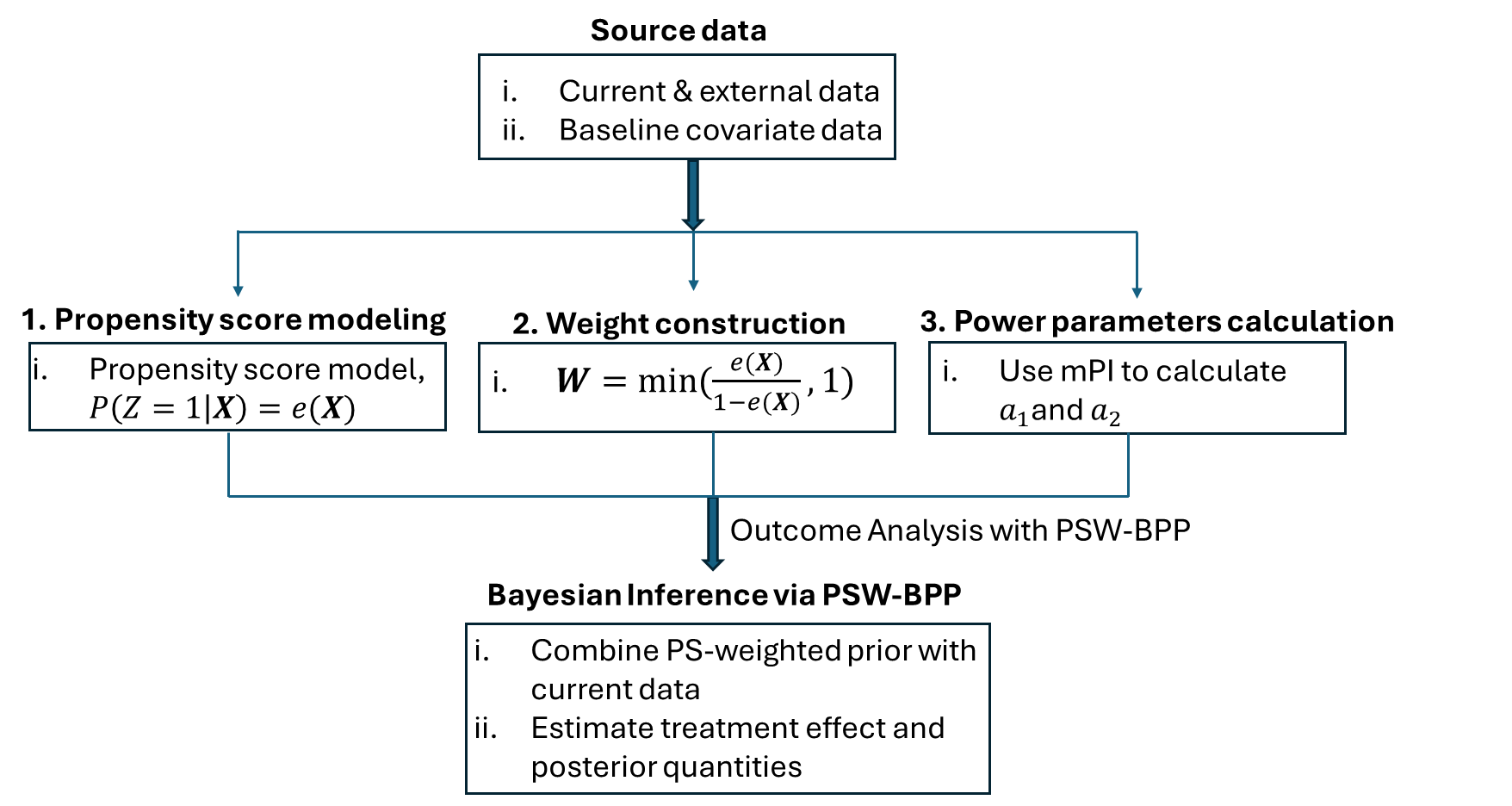}%
\caption{Illustration of the PSW-BPP approach}%
\label{fig:flowchart}%
\end{figure}

We will introduce our proposed method in this section. A flowchart of this proposed method is depicted in Figure \ref{fig:flowchart}. For the unknown treatment effect parameter $\theta_t$ and error variance $\sigma^2_t$, the likelihood of the current study treatment group $D_t$ can be written as 

\begin{equation}
    {\cal L} (\theta_t, \sigma_t^2 \mid D_t) 
    \propto (\sigma_t^2)^{-\frac{n_t}{2}} 
    \exp\left(-\frac{(\theta_t - \hat{\theta}_t)^\prime (\X_t^\prime \X_t)(\theta_t - \hat{\theta}_t)}{2\sigma_t^2}\right) \exp\left(-\frac{(n_t - 1) S_t^2}{2\sigma_t^2}\right),
\end{equation}
where $S_t^2=\frac{1}{n_t-1} (\Y_t-\X_t \hat \theta_t)^\prime (\Y_t-\X_t \hat \theta_t)$ and $\hat \theta_t=(\X_t^\prime \X_t)^{-1} \X_t^\prime \Y_t$. To complete the framework, we adopt the standard non-informative prior $\pi(\theta_t, \sigma^2_t) \propto 1/\sigma^2_t$. Combining this prior with the likelihood, the posterior distribution of $(\theta_t, \sigma^2_t)$ is given by

\begin{equation}
    \pi(\theta_t, \sigma_t^2 \mid D_t) \propto (\sigma_t^2)^{-\frac{n_t}{2}-1} 
    \exp\left(-\frac{(\theta_t - \hat{\theta}_t)^\prime (\X_t^\prime \X_t)(\theta_t - \hat{\theta}_t)}{2\sigma_t^2}\right) \exp\left(-\frac{(n_t - 1) S_t^2}{2\sigma_t^2}\right).
\end{equation}

From this joint posterior, the conditional distributions of the parameters can be directly derived. The coefficient of the treatment effect, $\theta_t \mid \sigma^2_t, D_t \sim {\cal N} \left(\hat \theta_t, \sigma^2_t (\X_t^\prime \X_t)^{-1} \right)$ and variance, $\sigma^2_t \mid D_t \sim \text{IG} \left(\frac{n_t-1}{2}, \frac{(n_t - 1) S_t^2}{2} \right)$. The details are presented in the section S1 of the supplementary materials.

We have a similar likelihood for the current study control group, which can be written as

\begin{equation}
    {\cal L} (\theta_c, \sigma_c^2 \mid D_c) 
    \propto (\sigma_c^2)^{-\frac{n_c}{2}} \exp\left(-\frac{(\theta_c - \hat{\theta}_c)^\prime (\X_c^\prime \X_c)(\theta_c - \hat{\theta}_c)}{2\sigma_c^2}\right)
    \exp\left(-\frac{(n_c - 1) S_c^2}{2\sigma_c^2}\right),
\end{equation}
where $S_c^2=\frac{1}{n_c-1} (\Y_c-\X_c \hat \theta_c)^\prime (\Y_c-\X_c \hat \theta_c)$.

The PSW-BPP can be written as
\begin{align}
    \pi(\theta_c, \sigma_c^2 \mid D_e, \ba) & \propto \left \{ (\sigma_c^2)^{-\frac{1}{2}} \exp\left(-\frac{(\theta_c - \hat{\theta}_e^w)^\prime (\X_e^\prime \W_e \X_e)(\theta_c - \hat{\theta}_e^w)}{2\sigma_c^2}\right) \right \}^{a_1} \\ \nonumber
    & \; \; \; \;\ \; \times \left \{ (S_e^{2,w})^{\frac{n_e^*-3}{2}} (\sigma^2_c)^{-\frac{n_e^*-1}{2}} \exp\left(-\frac{(n_e^* - 1) S_e^{2,w}}{2\sigma_c^2}\right) \right \}^{a_2} \pi_0(\theta_c, \sigma^2_c),
\end{align}
where $\W_e=\text{diag}(w_1,\cdots, w_{n_e})$ is a diagonal matrix of PS-derived weights, $\hat \theta_e^w=(\X_e^\prime\W_e\X_e)^{-1} \X_e^\prime \W_e \Y_e$, $S_e^{2,w}=\frac{1}{n_e^*-1} (\Y_e-\X_e \hat \theta_e^w)^\prime \W_e (\Y_e-\X_e \hat \theta_e^w)$, $\ba=(a_1, a_2)^\prime (0\leq a_1,a_2 \leq1)$ are the power parameters, and $n_e^*=\sum_{k=1}^{n_e} w_k$ is the effective sample size (ESS) in the external control group. Li et al. \cite{li2025using} proposed a weighting scheme for external subjects defined as

\begin{equation*}
    w_k=\frac{e(\X_{ek})}{1-e(\X_{ek})},
\end{equation*}
where $e(\X_{ek})$ denotes the estimated propensity score for the external subject. A direct application of this formula, however, can lead to weights that exceed 1, which may cause instability in subsequent analyses and reduce interpretability. To address this issue, we utilize a modified weighting rule by constraining it within the interval $[0,1]$, which is

\begin{equation}
    w_k=\min \left(1, \frac{e(\X_{ek})}{1-e(\X_{ek})}\right).
\end{equation}

The primary goal in this step is to assign balancing weights \cite{li2022estimands, li2018balancing} to every subject in the external control. It is crucial to adjust for differences in observed covariates between the two control groups, thereby mitigating potential confounding bias. This reweighting process helps create a more comparable study population, which in turn reduces the likelihood of prior–data conflict that can arise when the covariate distributions of the external and current study populations differ. Ultimately, the use of balancing weights improves the validity of the analysis by aligning the populations more closely and ensuring that any observed treatment effect is less influenced by systematic differences in baseline characteristics.

Using the likelihood and PSW-BPP, the joint posterior distribution of ($\theta_c, \sigma^2_c$) can be written as
\begin{equation}\begin{split}
\pi(\theta_c, \sigma_c^2 \mid D_c, D_e, \ba)
& \propto \exp\left\{-\frac{1}{2\sigma^2_c} \left[(\theta_c-\hat \theta_c)^\prime (\X_c^\prime \X_c) (\theta_c-\hat \theta_c) +a_1(\theta_c - \hat{\theta}_e^w)^\prime (\X_e^\prime \W_e \X_e)(\theta_c - \hat{\theta}_e^w)\right]\right\} \\
& \; \; \; \;\ \; \times (\sigma^2_c)^{-\frac{n_c+a_1}{2}} (S_e^{2,w})^{\frac{(n_e^*-3)a_2}{2}} (\sigma^2_c)^{-\frac{(n_e^*-1) a_2}{2}}\\ 
& \; \; \; \;\ \; \times \exp \left\{-\frac{1}{2\sigma^2_c} \left [(n_c-1)S^2_c + (n_e^*-1) a_2 S^{2,w}_e \right] \right \} \left(\frac{1}{\sigma^2_c}\right),
\end{split}\label{eq:control_posterior}
\end{equation}
where the initial prior $\pi_0(\theta_c, \sigma^2_c) \propto 1/\sigma^2_c$ is a non-informative prior.

Using some algebra, the conditional distributions of $\theta_c$ and $\sigma^2_c$ can be easily derived. The details are reported in the section S2 of the supplementary materials. The conditional posterior distribution of $\theta_c$ given others is

\begin{equation}
    \theta_c | \sigma_c^2, D_c, D_e, a_1 \sim {\cal N} (\tilde \theta_c, \Sigma_c),
\end{equation}
where
\begin{align*}
\tilde \theta_c & = (\X_c^\prime \X_c+a_1\X_e^\prime \W_e \X_e)^{-1} (\X_c^\prime \X_c\hat\theta_c+a_1\X_e^\prime \W_e \X_e\hat\theta_e^w), \\
\Sigma_c & = \sigma_c^2 (\X_c^\prime \X_c+a_1\X_e^\prime \W_e \X_e)^{-1}.
\end{align*}

Similarly, the conditional posterior distribution of $\sigma_c^2$ given others is
\begin{equation}
    \sigma_c^2|D_c, D_e,\ba \sim \mbox{IG} \left (\frac{n_c+ a_1+(n_e^*-1) a_2}{2}, \frac{(n_c-1)S_c^2+(n_e^*-1)a_2 S_e^{2,w}}{2} \right)
\end{equation}

\begin{remark}
    The posterior sampling of $\theta_c$ depends on $\sigma_c^2$, which is estimated using both current controls and the effective sample size contributed by external controls. When the external control group contains fewer than two subjects, $\sigma_c^2$ is computed solely from the current controls. This adjustment directly influences the calculation of the power parameters. To avoid unnecessary computational complexity in such cases, we set the power parameters to $a_1 = a_2 = 0$ whenever the external control group has fewer than two subjects.
\end{remark}

\subsection{Choice of Power Parameters} \label{sec:power_para}

Calculating the power parameters $\ba=(a_1, a_2)^\prime$ is a challenging task. One approach is to treat these parameters as random variables and assign them prior distributions, such as a Beta or a Uniform distribution. However, this strategy often introduces substantial mathematical and computational complexity. It is also possible to keep them fixed at different levels and perform a sensitivity analysis. As an alternative, Lu et al. \cite{lu2022propensity} proposed calculating the power parameters using an overlapping coefficient between the two data and a prespecified tuning parameter called nominal number of subjects to be borrowed from the external data.
To determine the overlapping coefficient, Inman and Bradley \cite{inman1989overlapping} proposed measuring the overlapping area of the propensity score density curves between the two studies. For additional details, readers are referred to Lu et al. \cite{lu2022propensity}. However, this approach requires solving a numerical integration problem between two density functions, which can be computationally demanding. Moreover, directly specifying the number of subjects to be borrowed might not be a reasonable assumption. 
Most importantly, these methods were developed for the calculation of stratum-specific power parameters and would not be directly applicable to our case. More recently, Baron et al. \cite{baron2024enhancing} introduced a new criterion, the minimal plausibility index (mPI), which comes with theoretical justification and provides optimal solutions for both $a_1 \ \text{and} \ a_2$.

The key idea is to measure how plausible it is that the external control population is similar to the current control population, and then set the borrowing strength accordingly. For the treatment effect, let $\delta=\theta_c-\theta_e$ denote the difference between the mean response in the current controls and external controls. The null hypothesis $H_0:\delta=0$ corresponds to the case where the two groups show exact similarity with respect to the mean, while the alternative hypothesis $H_1:\delta \neq 0$ reflects a difference. Under standard assumptions, the posterior distribution of $\delta$ follows a scaled $t$-distribution, $\frac{\bar Y_c - \bar Y_e}{S_c/\sqrt n_c} \mid (D_c, D_e) \sim t_{n_c-1}$, which is unimodal and symmetric about 0. The mPI for the mean is then defined as the posterior probability mass around $\delta=0$. A smaller probability suggests poor similarity, and the corresponding power parameter for borrowing the mean should be down-weighted.

For variance, define $\nu = \sigma_c^2 / \sigma_e^2$ as the ratio of variances between current and external controls. The null hypothesis $H_0:\nu = 1$ implies equal variability between the two groups, whereas the alternative $H_1: \nu \neq 1$ allows for differences in variance. Using the reference prior, the ratio of sample variances follows an $F$-distribution, $\frac{S^2_e}{S^2_c} \mid (D_c, D_e) \sim F_{n_e-1, n_c-1}$. The posterior distribution of $\nu$ is unimodal and the mPI for the variance corresponds to the posterior plausibility of $\nu=1$. When the posterior probability mass around $\nu = 1$ is small, the similarity in variance is low, and borrowing for the variance component should be reduced. The readers are suggested to review \cite{baron2024enhancing} for further details.

\section{Simulation Study} \label{sec:sim_study}

\subsection{Simulation Settings} \label{sec:sim_setting}
To assess the performance of our proposed methodology, we conduct extensive simulation studies. We assume that the current study was properly randomized, ensuring that the covariates in the treatment and current control groups are balanced, meaning they are from the same underlying distribution. We simulate independent covariates for all patients using a multivariate normal distribution with $p$ covariates, $\X_{ti}=(x_{ti1}, \dots, x_{tip}) \sim {\cal N}_p (\bmu_t, \bSigma)$, $\X_{cj}=(x_{cj1}, \dots, x_{cjp}) \sim {\cal N}_p (\bmu_c, \bSigma)$, and $\X_{ek}=(x_{ek1}, \dots, x_{ekp}) \sim {\cal N}_p (\bmu_e, \bSigma)$ for $i=1,\dots, n_t$, $j=1,\dots, n_c$, and $k=1,\dots, n_e$. The outcomes $Y_{ti}, Y_{cj}, \ \text{and} \ Y_{ek}$ are generated using the following equations

\begin{align*}
\Y_{ti} |\X_{ti} &= \beta_0 + \X'_{ti} \bbeta + \epsilon_{ti}, \\
\Y_{cj} |\x_{cj} &= \beta_0 + \X'_{cj} \bbeta + \epsilon_{cj}, \\
\Y_{ek} |\x_{ek} &= \beta_0 + \X'_{ek} \bbeta + \epsilon_{ek},
\end{align*}
where $\epsilon_{ti}, \epsilon_{cj}\sim {\cal N} (0,\sigma^2) \ \text{and} \ \epsilon_{ek} \sim {\cal N} (0,\eta^2)$ for $i=1,\cdots,n_t$, $j=1,\cdots,n_c$ and $k=1,\cdots,n_e$. We consider $n=n_t+n_c=100$ and 200, $n_e=1000$ and 2000, $p=4$, $\bSigma=I_4$, $\beta_0=0$, and $\bbeta=(\theta, 1,1.5,-1.3)$ where $\theta=2$ represents true treatment effect. 

We simulate the current study population under the assumption that $\bmu_t=\bmu_c=\bmu$. The first covariate, although continuous, is dichotomized at its median to define the treatment indicator: subjects with values below the median are assigned to the control group, while those with values above or equal to the median are assigned to the treatment group. In contrast, all $n_e$ subjects in the external study are assumed to belong to the control arm. It is important to note that the treatment variable is created by dichotomizing the first covariate; therefore, the original continuous version of this covariate is excluded from both the propensity score model and the outcome model. All remaining covariates, however, are included in both models. Our goal is to estimate the treatment effect of the current study while borrowing an optimal amount of information from an external source.

\subsection{Evaluation Measures} \label{sec:eva_mea}
We conducted simulation studies under twelve scenarios varying in covariate distributions and variance structures between the current and external control datasets to evaluate the performance of the proposed PSW-BPP method. All the scenarios are listed in Table \ref{tab:sim_scenario}. Each scenario is replicated 100 times for all simulation settings. We use 5,000 Gibbs samples after a burn-in of 2,000 iterations to compute all posterior estimates for each simulated dataset. We compute six posterior quantities for the overall treatment effects $\theta$ to evaluate the performance of the proposed model: 
(i) bias as $\text{Bias}=\frac{1}{B} \sum_{b=1}^{B} (\hat{\theta}_b - \theta_b)$ where $\hat{\theta}_b$ denotes the posterior mean of the $b$th replicated dataset and $\theta_b$ denotes the true parameter in the current group $b=1,\cdots,B$; 
(ii) absolute bias as $\text{ABias}= \frac{1}{B}\sum_{b=1}^{B} |\hat{\theta}_b - \theta_b|$; 
(iii) the root mean square error as $\text{RMSE}=\frac{1}{B} \sqrt{\sum_{b=1}^{B} (\hat{\theta}_b-\theta_b)^2}$; 
(iv) the simulation standard error as $\text{SE} = \sqrt{\frac{1}{B-1} \sum_{b=1}^{B} (\hat{\theta}_b - \frac{1}{B}\sum_{\ell=1}^{B} \hat{\theta}_\ell)^2}$; 
(v) the average width of 95\% credible intervals (Width); and 
(vi) the coverage probability (CP) of the 95\% credible intervals.

To compare the performance of our proposed method, we also evaluated an alternative method, which is based on the idea of PSW-BPP, but instead of using mPI to compute the power parameters, we fixed them at 0, 0.5, and 1. Setting $a_1=a_2=0$ implies that no information is borrowed. 
By comparing PSW-BPP with FB, we aim to evaluate how PS-weighting and data-driven optimal borrowing influence the accuracy and efficiency of the estimated treatment effect. 

\begin{table}[ht]
\caption{Different simulation settings. Note that the first element of each vector $``-"$ is related to the treatment variable}
\centering
\begin{tabular}{c c c c c}
\toprule
\textbf{Scenario} & \multicolumn{2}{c}{\textbf{Current study}} & \multicolumn{2}{c}{\textbf{External study}} \\ \cmidrule{2-5}
& $\mathbf{\bmu}$ & $\sigma^2$ & $\mathbf{\bmu}_e$ & $\eta^2$ \\ \bottomrule
I & (-, 1.2, 1.5, 1.6) & 1 & (-, 1, 1, 1) & 1 \\  
II & (-, 1, 1, 1) & 1 & (-, 1, 1, 1) & 1 \\ \hline

III & (-, 1.2, 1.5, 1.6) & 3 & (-, 1, 1, 1) & 3 \\  
IV & (-, 1, 1, 1) & 3 & (-, 1, 1, 1) & 3 \\ \hline

V & (-, 1.2, 1.5, 1.6) & 10 & (-, 1, 1, 1) & 10 \\ 
VI & (-, 1, 1, 1) & 10 & (-, 1, 1, 1) & 10 \\ \midrule

VII & (-, 1.2, 1.5, 1.6) & 1 & (-, 1, 1, 1) & 1.5 \\  
VIII & (-, 1, 1, 1) & 1 & (-, 1, 1, 1) & 1.5 \\ \hline

IX & (-, 1.2, 1.5, 1.6) & 3 & (-, 1, 1, 1) & 4 \\  
X & (-, 1, 1, 1) & 3 & (-, 1, 1, 1) & 4 \\ \hline

XI & (-, 1.2, 1.5, 1.6) & 10 & (-, 1, 1, 1) & 12 \\ 
XII & (-, 1, 1, 1) & 10 & (-, 1, 1, 1) & 12 \\ \bottomrule
\end{tabular}
\label{tab:sim_scenario}
\end{table}
\subsection{Simulation Results} \label{sec:sim_result}
While Table \ref{tab:mpi} presents the summaries of mPI used for optimal borrowing, Tables \ref{tab:simulation.treatment.100.equal}–\ref{tab:simulation.treatment.200.notequal} summarize the simulation results across all twelve scenarios. We can draw several conclusions from these results.

Table \ref{tab:mpi} shows that the mPI for the mean component remained relatively consistent (median around 0.25) across all twelve scenarios, while the variance-component mPI decreased sharply in heterogeneous cases (Scenarios IX–XII). This pattern confirms that PSW-BPP effectively adjusts the borrowing strength in response to heterogeneity between studies.

\begin{table}[ht]
\caption{Summaries of mPI over 100 iterations across various scenarios}
\centering
\begin{tabular}{c c c c c c c c c}
\toprule
& \multicolumn{4}{c}{$n = 100$} & \multicolumn{4}{c}{$n = 200$} \\ \cmidrule(lr){2-5} \cmidrule(lr){6-9}
Scenario & \multicolumn{2}{c}{mPI for mean} & \multicolumn{2}{c}{mPI for variance} 
         & \multicolumn{2}{c}{mPI for mean} & \multicolumn{2}{c}{mPI for variance} \\ \cmidrule(lr){2-3} \cmidrule(lr){4-5} \cmidrule(lr){6-7} \cmidrule(lr){8-9}
& Median & IQR & Median & IQR & Median & IQR & Median & IQR \\ \bottomrule
I   & 0.267  & 0.235  & 0.158  & 0.197  &  0.215 &  0.274 &  0.190 & 0.233  \\  
II  & 0.275 & 0.256 & 0.157 & 0.197 &  0.234 & 0.249  & 0.190  & 0.233  \\ \hline

III &  0.258 & 0.271  & 0.160  & 0.202 & 0.220  & 0.279  & 0.203  & 0.207  \\  
IV  & 0.280 & 0.268 & 0.160 & 0.202 & 0.207  &  0.219 & 0.203  &  0.207 \\ \hline

V   &  0.249 & 0.258 & 0.151  &  0.240 &  0.209 & 0.234  & 0.181  & 0.224  \\ 
VI  &  0.254 & 0.245  & 0.151  &  0.240 & 0.244  & 0.235  &  0.181 &  0.224 \\ \midrule

VII &  0.268 & 0.233  & 0.150  &  0.205 & 0.216  & 0.276  & 0.159  &  0.214 \\  
VIII&  0.278  & 0.256  & 0.150  & 0.205  &  0.233 & 0.248  &  0.159 & 0.214\\ \hline

IX  &  0.256 & 0.265  & 0.091 & 0.190  &  0.224 & 0.272  & 0.104  & 0.201  \\  
X   &  0.279 & 0.266  & 0.091  &  0.190 & 0.211  &  0.216 & 0.104  & 0.201  \\ \hline

XI  &  0.246 & 0.245  & 0.105  &  0.169 & 0.206  & 0.244  & 0.092  &  0.202 \\ 
XII &  0.251 &  0.246 & 0.105 & 0.169  &  0.246 & 0.231  & 0.092  &  0.202 \\ \bottomrule
\end{tabular}
\label{tab:mpi}
\end{table}

\begin{landscape}
\begin{table}[ht]
\centering
\caption{Simulation results for the overall treatment effect $\theta$ across scenarios I--VI ($n=100$, $n_e=1000$, and $\sigma^2=\eta^2$)}
\medskip
\begin{tabular}{ccccccccccccc} \toprule
Strategy     & Scenario    & $n$ & $n_e$  &   $\theta$     &  Mean      & Bias   & ABias  & RMSE       & SE    & Width & CP    \\ \toprule
PSW-BPP & I          & 100 & 1000 & 2 & 2.011 & $-0.011$ & 0.133 & 0.169 & 0.169 & 0.689 & 0.95 \\
FB $(a_1=a_2=0.0)$ & & 100 & 1000 & 2 & 2.013 & $-0.013$ & 0.148 & 0.189 & 0.189 & 0.738 & 0.94 \\
FB $(a_1=a_2=0.5)$ & & 100 & 1000 & 2 & 2.015 & $-0.015$ & 0.124 & 0.156 & 0.156 & 0.657 & 0.95 \\
FB $(a_1=a_2=1.0)$ & & 100 & 1000 & 2 & 2.016 & $-0.016$ & 0.118 & 0.149 & 0.149 & 0.625 & 0.96 \\  
 \hline

PSW-BPP & II         & 100 & 1000 & 2 & 2.015 & $-0.015$ & 0.123 & 0.157 & 0.157 & 0.662 & 0.96\\
FB $(a_1=a_2=0.0)$ & & 100 & 1000 & 2 & 2.015 & $-0.015$ & 0.137 & 0.177 & 0.177 & 0.704 & 0.93 \\
FB $(a_1=a_2=0.5)$ & & 100 & 1000 & 2 & 2.015 & $-0.015$ & 0.117 & 0.148 & 0.148 & 0.640 & 0.96 \\
FB $(a_1=a_2=1.0)$ & & 100 & 1000 & 2 & 2.015 & $-0.015$ & 0.112 & 0.140 & 0.140 & 0.614 & 0.97\\ 
 \hline  

PSW-BPP & III        & 100 & 1000 & 2 & 2.014 & $-0.014$ & 0.233 & 0.297 & 0.298 & 1.193 & 0.94 \\
FB $(a_1=a_2=0.0)$ & & 100 & 1000 & 2 & 2.023 & $-0.023$ & 0.256 & 0.327 & 0.328 & 1.278 & 0.94\\
FB $(a_1=a_2=0.5)$ & & 100 & 1000 & 2 & 2.026 & $-0.026$ & 0.214 & 0.270 & 0.271 & 1.138 & 0.95 \\
FB $(a_1=a_2=1.0)$ & & 100 & 1000 & 2 & 2.027 & $-0.027$ & 0.204 & 0.258 & 0.258 & 1.083 & 0.96 \\ 
 \hline  

PSW-BPP & IV         & 100 & 1000 & 2 & 2.023 & $-0.023$ & 0.217 & 0.277 & 0.277 & 1.147 & 0.95 \\
FB $(a_1=a_2=0.0)$ & & 100 & 1000 & 2 & 2.026 & $-0.026$ & 0.238 & 0.306 & 0.307 & 1.220 & 0.93 \\
FB $(a_1=a_2=0.5)$ & & 100 & 1000 & 2 & 2.027 & $-0.027$ & 0.203 & 0.256 & 0.256 & 1.108 & 0.96 \\
FB $(a_1=a_2=1.0)$ & & 100 & 1000 & 2 & 2.027 & $-0.027$ & 0.194 & 0.243 & 0.243 & 1.063 & 0.97\\ 
 \hline 

PSW-BPP & V          & 100 & 1000 &  2 & 2.023 & $-0.023$ & 0.434 & 0.553 & 0.555 & 2.179 & 0.94\\
FB $(a_1=a_2=0.0)$ & & 100 & 1000 &  2 & 2.041 & $-0.041$ & 0.467 & 0.599 & 0.599 & 2.334 & 0.94\\
FB $(a_1=a_2=0.5)$ & & 100 & 1000 & 2 & 2.047 & $-0.047$ & 0.391 & 0.494 & 0.494 & 2.078 & 0.95\\
FB $(a_1=a_2=1.0)$ & & 100 & 1000 & 2 & 2.050 & $-0.050$ & 0.373 & 0.472 & 0.471 & 1.977 & 0.96 \\ 
 \hline   

PSW-BPP & VI         & 100 & 1000 & 2 & 2.041 & $-0.041$ & 0.405 & 0.517 & 0.518 & 2.099 & 0.95 \\
FB $(a_1=a_2=0.0)$ & & 100 & 1000 & 2 & 2.048 & $-0.048$ & 0.434 & 0.560 & 0.560 & 2.227 & 0.93\\
FB $(a_1=a_2=0.5)$ & & 100 & 1000 & 2 & 2.049 & $-0.049$ & 0.371 & 0.467 & 0.467 & 2.024 & 0.96 \\
FB $(a_1=a_2=1.0)$ & & 100 & 1000 & 2 & 2.049 & $-0.049$ & 0.354 & 0.044 & 0.044 & 1.941 & 0.97 \\ 
\bottomrule
\end{tabular}
\label{tab:simulation.treatment.100.equal}
\end{table}
\end{landscape}
\begin{landscape}
\begin{table}[ht]
\centering
\caption{Simulation results for the overall treatment effect $\theta$ across scenarios VII--XII ($n=100$, $n_e=1000$, and $\sigma^2 \neq \eta^2$)}
\medskip
\begin{tabular}{ccccccccccccc} \toprule
Strategy     & Scenario    & $n$ & $n_e$  &   $\theta$     &  Mean      & Bias   & ABias  & RMSE       & SE    & Width & CP    \\ \toprule
PSW-BPP & VII        & 100 & 1000 & 2 & 2.011 & $-0.011$ & 0.134 & 0.170 & 0.170 & 0.710 & 0.95 \\
FB $(a_1=a_2=0.0)$ & & 100 & 1000 & 2 & 2.013 & $-0.013$ & 0.148 & 0.189 & 0.189 & 0.738 & 0.94 \\
FB $(a_1=a_2=0.5)$ & & 100 & 1000 & 2 & 2.015 & $-0.015$ & 0.125 & 0.158 & 0.158 & 0.711 & 0.97 \\
FB $(a_1=a_2=1.0)$ & & 100 & 1000 & 2 & 2.016 & $-0.016$ & 0.120 & 0.152 & 0.152 & 0.700 & 0.97 \\ 
 \hline

PSW-BPP & VIII       & 100 & 1000 & 2 & 2.015 & $-0.015$ & 0.124 & 0.157 & 0.157 & 0.682 & 0.96 \\
FB $(a_1=a_2=0.0)$ & & 100 & 1000 & 2 & 2.015 & $-0.015$ & 0.137 & 0.177 & 0.177 & 0.704 & 0.93 \\
FB $(a_1=a_2=0.5)$ & & 100 & 1000 & 2 & 2.015 & $-0.015$ & 0.118 & 0.148 & 0.148 & 0.692 & 0.98 \\
FB $(a_1=a_2=1.0)$ & & 100 & 1000 & 2 & 2.015 & $-0.015$ & 0.113 & 0.141 & 0.141 & 0.687 & 0.98 \\ 
 \hline 

PSW-BPP & IX         & 100 & 1000 & 2 & 2.014 & $-0.014$ & 0.233 & 0.298 & 0.300 & 1.213 & 0.93 \\
FB $(a_1=a_2=0.0)$ & & 100 & 1000 & 2 & 2.023 & $-0.023$ & 0.256 & 0.327 & 0.328 & 1.278 & 0.94 \\
FB $(a_1=a_2=0.5)$ & & 100 & 1000 & 2 & 2.026 & $-0.026$ & 0.216 & 0.273 & 0.273 & 1.200 & 0.96 \\
FB $(a_1=a_2=1.0)$ & & 100 & 1000 & 2 & 2.028 & $-0.028$ & 0.206 & 0.262 & 0.262 & 1.171 & 0.97 \\ 
 \hline 

PSW-BPP & X          & 100 & 1000 & 2 & 2.023 & $-0.023$ & 0.217 & 0.277 & 0.278  & 1.166 & 0.96 \\
FB $(a_1=a_2=0.0)$ & & 100 & 1000 & 2 & 2.026 & $-0.026$ & 0.238 & 0.306 & 0.307 & 1.220 & 0.93\\
FB $(a_1=a_2=0.5)$ & & 100 & 1000 &  2 & 2.026 & $-0.027$ & 0.204 & 0.256 & 0.256 & 1.170 & 0.98\\
FB $(a_1=a_2=1.0)$ & & 100 & 1000 & 2 & 2.027 & $-0.027$ & 0.195 & 0.244 & 0.244 & 1.149 & 0.98 \\ 
 \hline

PSW-BPP & XI         & 100 & 1000 & 2 & 2.023 & $-0.023$ & 0.435 & 0.554 & 0.556 & 2.196 & 0.94 \\
FB $(a_1=a_2=0.0)$ & & 100 & 1000 & 2 & 2.041 & $-0.041$ & 0.467 & 0.597 & 0.599 & 2.334 & 0.94 \\
FB $(a_1=a_2=0.5)$ & & 100 & 1000 & 2 & 2.048 & $-0.048$ & 0.393 & 0.496 & 0.496 & 2.147 & 0.96 \\
FB $(a_1=a_2=1.0)$ & & 100 & 1000 & 2 & 2.050 & $-0.050$ & 0.375 & 0.475 & 0.475 & 2.075 & 0.97 \\ 
 \hline  

PSW-BPP & XII        & 100 & 1000 & 2 & 2.041 & $-0.041$ & 0.405 & 0.518 & 0.519 & 2.116 & 0.95 \\
FB $(a_1=a_2=0.0)$ & & 100 & 1000 & 2 & 2.048 & $-0.048$ & 0.434 & 0.560 & 0.560 & 2.227 & 0.93 \\
FB $(a_1=a_2=0.5)$ & & 100 & 1000 & 2 & 2.048 & $-0.048$ & 0.372 & 0.468 & 0.468 & 2.091 & 0.97 \\
FB $(a_1=a_2=1.0)$ & & 100 & 1000 & 2 & 2.049 & $-0.049$ & 0.355 & 0.445 & 0.445 & 2.037 & 0.98 \\ 
 \bottomrule
\end{tabular}
\label{tab:simulation.treatment.100.notequal}
\end{table}
\end{landscape}
\begin{landscape}
\begin{table}[ht]
\centering
\caption{Simulation results for the overall treatment effect $\theta$ across scenarios I--VI ($n=200$, $n_e=2000$, and $\sigma^2=\eta^2$)}
\medskip
\begin{tabular}{ccccccccccccc} \toprule
Strategy     & Scenario    & $n$ & $n_e$  &   $\theta$     &  Mean      & Bias   & ABias  & RMSE       & SE    & Width & CP    \\ \toprule
PSW-BPP & I          & 200 & 2000 & 2 & 1.999 & 0.001 & 0.090 & 0.113 & 0.114 & 0.492 & 0.95 \\
FB $(a_1=a_2=0.0)$ & & 200 & 2000 & 2 & 2.006 & $-0.006$ & 0.099 & 0.124 & 0.125 & 0.524 & 0.94 \\
FB $(a_1=a_2=0.5)$ & & 200 & 2000 & 2 & 2.002 & $-0.002$  & 0.085 & 0.105 & 0.105 & 0.465 & 0.97\\
FB $(a_1=a_2=1.0)$ & & 200 & 2000 & 2 & 2.001 & $-0.001$ & 0.084 & 0.102 & 0.103 & 0.442 & 0.98 \\ 
 \hline

PSW-BPP & II         & 200 & 2000 & 2 & 2.007 & $-0.007$ & 0.090 & 0.111 & 0.112 & 0.474 & 0.96\\
FB $(a_1=a_2=0.0)$ & & 200 & 2000 & 2 & 2.007 & $-0.007$ & 0.097 & 0.120 & 0.121 & 0.500 & 0.97 \\
FB $(a_1=a_2=0.5)$ & & 200 & 2000 & 2 & 2.004 & $-0.004$ & 0.083 & 0.104 & 0.104 & 0.454 & 0.97 \\
FB $(a_1=a_2=1.0)$ & & 200 & 2000 & 2 & 2.003 & $-0.003$ & 0.080 & 0.101 & 0.101 & 0.435 & 0.97 \\ 
 \hline 

PSW-BPP & III        & 200 & 2000 & 2 & 1.998 & 0.002 & 0.158 & 0.198 & 0.199 & 0.851 & 0.95 \\
FB $(a_1=a_2=0.0)$ & & 200 & 2000 & 2 & 2.011 & $-0.011$ & 0.173 & 0.215 & 0.216 & 0.907 & 0.94\\
FB $(a_1=a_2=0.5)$ & & 200 & 2000 & 2 & 2.003 & $-0.003$ & 0.147 & 0.181 & 0.182 & 0.805 & 0.97\\
FB $(a_1=a_2=1.0)$ & & 200 & 2000 & 2 & 2.001 & $-0.001$ & 0.145 & 0.177 & 0.178 & 0.765 & 0.98\\ 
 \hline 

PSW-BPP & IV         & 200 & 2000 & 2 & 2.011 & $-0.011$ & 0.156 & 0.194 & 0.195 & 0.822 & 0.96 \\
FB $(a_1=a_2=0.0)$ & & 200 & 2000 & 2 & 2.012 & $-0.012$ & 0.168 & 0.208 & 0.209 & 0.866 & 0.97 \\
FB $(a_1=a_2=0.5)$ & & 200 & 2000 & 2 & 2.007 & $-0.007$ & 0.144 & 0.179 & 0.180 & 0.789 & 0.97 \\
FB $(a_1=a_2=1.0)$ & & 200 & 2000 & 2 & 2.005 & $-0.005$ & 0.139 & 0.174 & 0.175 & 0.754 & 0.97 \\ 
 \hline 

PSW-BPP & V          & 200 & 2000 & 2 & 1.998 & 0.002 & 0.291 & 0.365 & 0.367 & 1.553 & 0.95\\
FB $(a_1=a_2=0.0)$ & & 200 & 2000 & 2 & 2.019 & $-0.019$ & 0.316 & 0.392 & 0.394 & 1.655 & 0.94 \\
FB $(a_1=a_2=0.5)$ & & 200 & 2000 & 2 & 2.006 & $-0.006$ & 0.268 & 0.331 & 0.332 & 1.470 & 0.97 \\
FB $(a_1=a_2=1.0)$ & & 200 & 2000 & 2 & 2.001 & $-0.001$ & 0.264 & 0.323 & 0.325 & 1.397 & 0.98 \\ 
 \hline   

PSW-BPP & VI         & 200 & 2000 & 2 & 2.017 & $-0.017$ & 0.289 & 0.355 & 0.357 & 1.500 & 0.96 \\
FB $(a_1=a_2=0.0)$ & & 200 & 2000 & 2 & 2.021 & $-0.021$ & 0.307 & 0.380 & 0.381 & 1.581 & 0.97 \\
FB $(a_1=a_2=0.5)$ & & 200 & 2000 & 2 & 2.012 & $-0.012$ & 0.263 & 0.327 & 0.329 & 1.435 & 0.97 \\
FB $(a_1=a_2=1.0)$ & & 200 & 2000 & 2 & 2.001 & $-0.001$ & 0.254 & 0.318 & 0.319 & 1.377 & 0.97 \\ 
\bottomrule
\end{tabular}
\label{tab:simulation.treatment.200.equal}
\end{table}
\end{landscape}
\begin{landscape}
\begin{table}[ht]
\centering
\caption{Simulation results for the overall treatment effect $\theta$ across scenarios VII--XII ($n=200$, $n_e=2000$, and $\sigma^2 \neq \eta^2$)}
\medskip
\begin{tabular}{ccccccccccccc} \toprule
Strategy     & Scenario    & $n$ & $n_e$  &   $\theta$     &  Mean      & Bias   & ABias  & RMSE       & SE    & Width & CP    \\ \toprule
PSW-BPP & VII         & 200 & 2000 & 2 & 1.999 & 0.001 & 0.091 & 0.114 & 0.114 & 0.508 & 0.97 \\
FB $(a_1=a_2=0.0)$  & & 200 & 2000 & 2 & 2.006 & $-0.006$ & 0.100 & 0.124 & 0.125 & 0.523 & 0.94 \\
FB $(a_1=a_2=0.5)$  & & 200 & 2000 & 2 & 2.001 & $-0.001$ & 0.085 & 0.105 & 0.106 & 0.502 & 0.98 \\
FB $(a_1=a_2=1.0)$  & & 200 & 2000 & 2 & 1.999 & 0.001 & 0.085 & 0.104 & 0.104 & 0.494 & 0.98 \\ 
 \hline

PSW-BPP & VIII         & 200 & 2000 & 2 & 2.007 & $-0.007$ & 0.090 & 0.111 & 0.112 & 0.490 & 0.96 \\
FB $(a_1=a_2=0.0)$   & & 200 & 2000 & 2 & 2.007 & $-0.007$ & 0.097 & 0.121 & 0.121 & 0.500 & 0.97 \\
FB $(a_1=a_2=0.5)$   & & 200 & 2000 & 2 & 2.004 & $-0.004$ & 0.083 & 0.104 & 0.104 & 0.490 & 0.98 \\
FB $(a_1=a_2=1.0)$   & & 200 & 2000 & 2 & 2.002 & $-0.002$ & 0.081 & 0.101 & 101 & 0.487 & 0.98 \\ 
 \hline  

PSW-BPP & IX         & 200 & 2000 & 2 & 1.998 & 0.002 & 0.159 & 0.199 & 0.200 & 00.867 & 0.97 \\
FB $(a_1=a_2=0.0)$ & & 200 & 2000 & 2 & 2.011 & $-0.011$ & 0.173 & 0.215 & 0.216 & 0.907 & 0.94 \\
FB $(a_1=a_2=0.5)$ & & 200 & 2000 & 2 & 2.003 & $-0.003$ & 0.148 & 0.182 & 0.183 & 0.848 & 0.98 \\
FB $(a_1=a_2=1.0)$ & & 200 & 2000 & 2 & 1.999 & 0.001 & 0.147 & 0.179 & 0.180 & 0.827 & 0.98 \\ 
 \hline  

PSW-BPP & X          & 200 & 2000 & 2 & 2.011 & $-0.011$ & 0.158 & 0.194 & 0.195 & 0.837 & 0.96 \\
FB $(a_1=a_2=0.0)$ & & 200 & 2000 & 2 & 2.012 & $-0.012$ & 0.168 & 0.208 & 0.209 & 0.866 & 0.97 \\
FB $(a_1=a_2=0.5)$ & & 200 & 2000 & 2 & 2.006 & $-0.006$ & 0.144 & 0.180 & 0.180 & 0.829 & 0.98 \\
FB $(a_1=a_2=1.0)$ & & 200 & 2000 & 2 & 2.004 & $-0.004$ & 0.140 & 0.175 & 0.175 & 0.814 & 0.98 \\ 
 \hline

PSW-BPP & XI         & 200 & 2000 & 2 & 1.997 & 0.003 & 0.292 & 0.365 & 0.367 & 1.566 & 0.96 \\
FB $(a_1=a_2=0.0)$ & & 200 & 2000 & 2 & 2.019 & $-0.019$ & 0.316 & 0.392 & 0.394 & 1.655 & 0.94 \\
FB $(a_1=a_2=0.5)$ & & 200 & 2000 & 2 & 2.005 & $-0.005$ & 0.269 & 0.332 & 0.333 & 1.518 & 0.98 \\
FB $(a_1=a_2=1.0)$ & & 200 & 2000 & 2 & 2.001 & $-0.001$ & 0.266 & 0.325 & 0.327 & 1.465 & 0.98 \\ 
 \hline  

PSW-BPP & XII        & 200 & 2000 & 2 & 2.017 & $-0.017$ & 0.289 & 0.355 & 0.357 & 1.513 & 0.96\\
FB $(a_1=a_2=0.0)$ & & 200 & 2000 & 2 & 2.021 & $-0.021$ & 0.307 & 0.380 & 0.381 & 1.581 & 0.97 \\
FB $(a_1=a_2=0.5)$ & & 200 & 2000 & 2 & 2.012 & $-0.012$ & 0.264 & 0.328 & 0.329 & 1.482 & 0.97 \\
FB $(a_1=a_2=1.0)$ & & 200 & 2000 & 2 & 2.008 & $-0.008$ & 0.255 & 0.318 & 0.320 & 1.444 & 0.98 \\ 
  \bottomrule
\end{tabular}
\label{tab:simulation.treatment.200.notequal}
\end{table}
\end{landscape}

Tables \ref{tab:simulation.treatment.100.equal} and \ref{tab:simulation.treatment.200.equal} present simulation scenarios I-VI for varying sample sizes where variances are homogeneous ($\sigma^2=\eta^2$) between the two studies. For these settings,  PSW-BPP produced nearly unbiased estimates of the overall treatment effect $\theta$ across all scenarios. Bias and RMSE remained small, and 95\% credible intervals achieved nominal coverage levels. Compared with the FB method when no information is leveraged ($a_1=a_2=0$), PSW-BPP yielded a smaller RMSE and narrower interval widths, while maintaining comparable or slightly higher coverage probabilities. However, when more information is leveraged ($a_1=a_2 \geq 0.5$), FB produced lower posterior quantities for all scenarios, which might be due to overborrowing. This suggests that fixed borrowing may lead to excessive influence from external data when the current and external studies differ, whereas the adaptive nature of PSW-BPP effectively controls the borrowing strength and yields more reliable inference. Second, the results indicate that when heterogeneity exist between the external and current studies, as observed in scenarios I, III, and V, the values of ABias, RMSE, SE, and interval width increase compared to their corresponding homogeneous scenarios (II, IV, and VI). This finding suggests that borrowing information is more effective when the two studies are more comparable. Third, as the current study sample size $n$ increases from 100 to 200  and the external study sample size $n_e$ increases from 1000 to 2000, all methods exhibited lower ABias, RMSE, and SE, along with narrower interval widths. PSW-BPP continued to demonstrate consistent estimation accuracy and reliable uncertainty quantification across sample sizes, confirming its scalability when larger external datasets are available for borrowing. 
These findings suggest that incorporating external information through the BPP leads to estimates that are more accurate and more consistent than those obtained without borrowing. In particular, the improved coverage stability indicates that controlled borrowing can enhance the reliability of inference by reducing variability and mitigating bias relative to the no-borrowing approach. Finally, the data-driven mPI mechanism and propensity score weighting allow PSW-BPP to dynamically determine borrowing strength rather than relying on a fixed nominal borrowed size, yielding more robust and adaptive inference.

Tables \ref{tab:simulation.treatment.100.notequal} and \ref{tab:simulation.treatment.200.notequal} present results when the variance between two datasets is heterogeneous ($\sigma^2\neq\eta^2$; Scenarios VII–XII). PSW-BPP automatically reduced borrowing from the external variance component, as indicated by lower variance–mPI values in Table \ref{tab:mpi}. This adaptive down-weighting prevented overly narrow credible intervals and maintained coverage near the nominal 95\% level. Compared to FB, PSW-BPP provided stable inference with modest interval widths and minimal bias, highlighting its robustness against heterogeneous external information.

Overall, these results demonstrate that PSW-BPP efficiently incorporates external information when the current and external controls are homogeneous and automatically reduces borrowing strength when heterogeneity arise. The proposed approach provides accurate, efficient, and robust estimates of the treatment effect $\theta$ across a wide range of data-generating scenarios and sample sizes.


\section{Case Study} \label{sec:case_study}
Although the proposed method is designed to make clinical trials more ethical and efficient by reducing the number of patients required, we were unable to identify suitable subject-level data in the public domain. Therefore, we demonstrate the application of the proposed methodology by reformatting data from a retrospective, non-randomized study to mimic a RCT setting. We use the Alzheimer’s Disease Neuroimaging Initiative (ADNI), a well-established database for Alzheimer’s disease research, from which we construct a quasi-randomized exposure–control trial and an external control arm. While this does not represent a perfect example of case study for the proposed method, it adequately serves the purpose of methodological demonstration.

Considered the gold standard for evaluating the efficacy of anti-dementia treatments \cite{kueper2018alzheimer}, the ADAS13 consists of a sequence of 13 cognitive tasks, with the total score defined as the sum of errors across tasks; higher scores indicate worse cognitive performance. We define the exposure effect size as the difference between baseline and week 52 ADAS13 scores, where a positive difference corresponds to an improvement in cognitive function for a given subject. The baseline covariates included in the analysis are age, gender, RAVLT (Rey Auditory Verbal Learning Test), APOE4 status, and MMSE (Mini-Mental State Examination) score. Although five baseline diagnostic categories were available, we restricted ourselves to the three most relevant, cognitively normal (CN), late mild cognitive impairment (LMCI), and Alzheimer’s disease (AD) to define the exposure–control variable. Specifically, CN serves as the control group, while LMCI and AD together form the exposure group. APOE4 status is categorized as 0, 1, or 2 alleles, and MMSE scores range from 18 to 30. The RAVLT measure comprises three subcomponents—RAVLT immediate, learning, and forgetting—which we average to construct a single composite RAVLT variable. The primary dataset includes three cohorts defined by enrollment period. We treat ADNI1 as external data and ADNI2 and ADNI3 as current data, yielding 531 subjects in the current study and 710 subjects in the external study cohort. The primary objective of this analysis is to estimate the exposure effect size while borrowing information from the external data. A summary of the baseline covariates can be obtained in Table \ref{tab:base.summ}.

\begin{table}[ht]
\centering
\caption{Baseline characteristics of the current and external study population}
\medskip
\begin{tabular}{llll} \hline
\textbf{Characteristics}    & \textbf{Current study} & \textbf{External study} & \textbf{$p$-value}\\ 
                            &$n=531$                            &$n=710$                 \\ \hline
Exposure (\%)               &  63.65 & 70.99 & 0.007\\  
Gender Male (\%)            & 55.18  & 57.75 & 0.398\\
AGE (mean/SD)               & 73.19 (7.36) & 75.27 (6.73) & $<0.001$ \\ 
Baseline RAVLT (mean/SD)    & 14.42 (5.03) & 13.73 (4.44) & 0.012 \\ 
APOE4 $0$ (/\%)             & 50.65 & 51.41 & 0.500\\ 
\hspace{10.5mm}  $1$ (/\%)  & 36.34 & 37.75 & \\ 
Baseline MMSE  (mean/SD)    & 26.95 (2.93) & 26.88 (2.58) & 0.629 \\
\hline
\end{tabular}
\label{tab:base.summ}
\vspace{0.5em}
\footnotesize
Note that $p$-values were calculated using Chi-squared test for categorical variables and the two-sample $t$-test for continuous variables.
\end{table}

We implement the proposed PSW-BPP method to estimate the exposure effect size. The hypotheses are specified as $H_0: \theta \leq 0 \ \text{vs.} \ H_a:\theta > 0$, where $\theta$ represents the exposure effect, defined as the difference in mean change in ADAS13 scores between the exposure and control groups. A larger positive $\theta$ corresponds to greater cognitive improvement. It is assumed that the null hypothesis is rejected if the posterior probability of $\theta > 0$ exceeds the prespecified threshold of 0.975.

PS are estimated using a logistic regression model, and subject-specific weights are assigned to individuals in the external cohort. These weights are incorporated into the BPP framework, allowing external data to contribute information in a manner proportional to their similarity to the current study. Using the PSW-BPP approach, the estimated exposure effect is $\hat{\theta} = 0.166$, with a 95\% credible interval of $(0.039, 0.309)$, indicating a statistically significant positive effect. This result suggests that exposure is associated with reduced ADAS13 scores, reflecting improved cognitive performance.

The posterior probability that $\theta > 0$ is 99.3\%, which meets the predefined study success criterion. The effective sample size contributed by the external data under this framework is 153.66, reflecting the partial borrowing enabled by the PSW-BPP method. Focusing only on the current study without borrowing external data yields $\hat{\theta} = 0.175$ with a 95\% credible interval of $(0.039, 0.326)$, demonstrating that incorporating external data does not meaningfully alter the point estimate. However, borrowing reduces the width of the credible interval, providing a more precise and robust estimate of the exposure effect. The two power parameters are estimated as $a_1=0.482 \ \text{and} \ a_2=0.215$. Hence, more information is borrowed from the mean part than the variance.

Overall, the PSW-BPP approach allows for the ethical and efficient incorporation of external information while maintaining the integrity of the inference for the current study.

\section{Discussion} \label{sec:discussion}
In this paper, we introduce a unified framework, PSW-BPP, to augment the control arm of an RCT by borrowing information from external control data. The proposed method integrates propensity score weighting with a BPP, employing data-adaptive power parameters that separately regulate the amount of information borrowed for the mean and variance components of the outcome model. This integration enables flexible, component-wise borrowing that reflects the degree of compatibility between the current and external control data, while preserving valid and efficient inference for the treatment effect of interest.

The proposed method offers several methodological advantages over the approach of Li et al. \cite{li2025using}. First, we modify the PS weighting scheme to ensure that the resulting weights are constrained within the interval  $[0,1]$, which improves numerical stability and prevents undue influence from individual external observations. Second, rather than adopting a traditional power prior with a single global power parameter, we employ a BPP with two distinct power parameters that enable selective and component-wise borrowing of information. This formulation is particularly advantageous in settings where the mean and variance components of the outcome model exhibit differential levels of heterogeneity between the current and external data sources. Third, we leverage the concept of the mPI to guide the calibration of the power parameters, providing a data-driven mechanism that mitigates the risks of over-borrowing or under-borrowing and enhances the robustness of the resulting inference.

Simulation studies demonstrate several key advantages of PSW-BPP over competing borrowing methods. Across a broad range of data-generating scenarios, including varying covariate distributions, residual variances, and sample sizes, PSW-BPP consistently yielded nearly unbiased estimates of the overall treatment effect with appropriate uncertainty quantification. Compared with fixed-borrowing approaches, PSW-BPP achieved lower RMSE and produced narrower credible intervals. Another key strength of PSW-BPP lies in its ability to adaptively down-weight external information when discrepancies between studies are present.  When heterogeneity in variance or covariate distributions between the current and external controls was present, the mPI-driven borrowing mechanism automatically reduced the degree of borrowing, thereby preventing overly optimistic uncertainty estimates. This adaptive behavior contrasts with traditional fixed power priors and related approaches, which may continue to borrow aggressively despite partial incompatibility between data sources, potentially leading to undercoverage \cite{ibrahim2000power, hobbs2011hierarchical}.

Comparisons with alternative borrowing strategies further highlight the advantages of PSW-BPP. Relative to fixed-borrowing priors with prespecified power parameters, PSW-BPP eliminates the need for subjective tuning and reduces sensitivity to prior misspecification. Unlike stratification-based methods \cite{wang2019propensity, lu2022propensity, baron2024enhancing} that rely on a fixed nominal number of sample size to be borrowed, PSW-BPP allows borrowing to vary flexibly across model components in a fully data-driven manner. In nearly all simulation settings, the proposed method achieved comparable or improved efficiency while maintaining more stable coverage. Moreover, PSW-BPP scaled effectively with increasing sample sizes, continuing to deliver accurate point estimates and well-calibrated intervals as both current and external datasets grew. This scalability is particularly relevant in contemporary clinical research, where large external data sources, including real-world evidence, are increasingly leveraged \cite{hernan2016using}.

Although PSW-BPP offers several methodological advantages, it also has a few limitations. First, the proposed framework assumes that all relevant confounders are observed and adequately captured in the PS model. As with all PS–based methods, unmeasured confounding may compromise the validity of the weighted likelihood and, consequently, the borrowing mechanism \cite{rosenbaum1983central, stuart2010matching}. Second, the current implementation focuses on continuous outcomes with Gaussian errors; extensions to generalized linear models and time-to-event outcomes would broaden the method's applicability. Third, while the mPI provides an interpretable and effective measure of commensurability, alternative discrepancy metrics or hierarchical extensions could be explored to further enhance robustness.

Future research may extend PSW-BPP to more complex data settings, including binary and time-to-event outcomes, longitudinal responses, and multilevel or clustered data structures. Although the proposed methodology is developed for a single external data source, it can be naturally extended to accommodate multiple external datasets by allowing dataset-specific weighting and borrowing parameters. In addition, incorporating flexible machine learning–based PS models may further improve covariate balance in high-dimensional settings. Additionally, a variable selection can be performed within the PS model for high-dimensional data.  Finally, a deeper investigation of the theoretical properties of the mPI, particularly under model misspecification, would provide valuable insight into the robustness and operating characteristics of the proposed borrowing mechanism.



\backmatter

\section*{Supplementary Information}
Online Supplementary Materials include two sections: 
(i) S1 (Inference for treatment arm) and (ii) S2 (Inference for control arm).

\section*{Acknowledgements} Data collection and sharing for this project was funded by the Alzheimer's Disease Neuroimaging Initiative (ADNI) (National Institutes of Health Grant U01 AG024904) and DOD ADNI (Department of Defense award number W81XWH-12-2-0012). ADNI is funded by the National Institute on Aging, the National Institute of Biomedical Imaging and Bioengineering, and through generous contributions from the following: AbbVie, Alzheimer’s Association; Alzheimer’s Drug Discovery Foundation; Araclon Biotech; BioClinica, Inc.; Biogen; Bristol-Myers Squibb Company; CereSpir, Inc.; Cogstate; Eisai Inc.; Elan Pharmaceuticals, Inc.; Eli Lilly and Company; EuroImmun; F. Hoffmann-La Roche Ltd and its affiliated company Genentech, Inc.; Fujirebio; GE Healthcare; IXICO Ltd.; Janssen Alzheimer Immunotherapy Research \& Development, LLC.; Johnson \& Johnson Pharmaceutical Research \& Development LLC.; Lumosity; Lundbeck; Merck \& Co., Inc.; Meso Scale Diagnostics, LLC.; NeuroRx Research; Neurotrack Technologies; Novartis Pharmaceuticals Corporation; Pfizer Inc.; Piramal Imaging; Servier; Takeda Pharmaceutical Company; and Transition Therapeutics. The Canadian Institutes of Health Research is providing funds to support ADNI clinical sites in Canada. Private sector contributions are facilitated by the Foundation for the National Institutes of Health (\url{www.fnih.org}). The grantee organization is the Northern California Institute for Research and Education, and the study is coordinated by the Alzheimer’s Therapeutic Research Institute at the University of Southern California. ADNI data are disseminated by the Laboratory for Neuro Imaging at the University of Southern California.

\section*{Declarations}

\begin{itemize}
\item Funding: Not applicable.
\item Conflict of interest/Competing interests: Authors declare no potential conflict of interest.
\item Ethics approval and consent to participate: Not applicable.
\item Data availability: Data can be obtained from \href{https://adni.loni.usc.edu/}{ADNI} upon request.
\item Code availability: All code is written in statistical software R 4.2.2 and the code
can be downloaded from the \href{https://github.com/apustat/PSW-BPP}{GitHub}.
\end{itemize}

\begin{appendices}





\end{appendices}


\bibliography{sn-bibliography}

\end{document}